\documentclass[sigconf]{acmart}

\usepackage{balance}
\usepackage{booktabs}
\usepackage{color}
\usepackage{tikz}
\usepackage{enumerate}
\usepackage{hyperref}
\usepackage{cleveref}
\usepackage{esint}

\usepackage{subfigure}
\usepackage{algorithm}
\usepackage{algorithmic}
\usepackage{multirow}
\def\BibTeX{{\rm B\kern-.05em{\sc i\kern-.025em b}\kern-.08emT\kern-.1667em\lower.7ex\hbox{E}\kern-.125emX}}

\begin{document}

\copyrightyear{2019}
\acmYear{2019}
\setcopyright{acmcopyright}
\acmConference[KDD '19]{The 25th ACM SIGKDD Conference on Knowledge Discovery and Data Mining}{August 4--8, 2019}{Anchorage, AK, USA}
\acmPrice{15.00}
\acmDOI{10.1145/3292500.3330882}
\acmISBN{978-1-4503-6201-6/19/08}

\title{EdMot: An Edge Enhancement Approach for Motif-aware Community Detection}

\author{Pei-Zhen Li, Ling Huang, Chang-Dong Wang, Jian-Huang Lai}
\affiliation{%
  \institution{School of Data and Computer Science, Sun Yat-sen University, Guangzhou, China\\
  Guangdong Province Key Laboratory of Computational Science, Guangzhou, China\\
   Key Laboratory of Machine Intelligence and Advanced Computing, Ministry of Education, China}
}
\email{sysuLiPeizhen@163.com,huanglinghl@hotmail.com,changdongwang@hotmail.com,stsljh@mail.sysu.edu.cn}

%

\begin{abstract}
Network community detection is a hot research topic in network analysis. Although many methods have been proposed for community detection, most of them only take into consideration the lower-order structure of the network at the level of individual nodes and edges. Thus, they fail to capture the higher-order characteristics at the level of small dense subgraph patterns, e.g., motifs. Recently, some higher-order methods have been developed but they typically focus on the motif-based hypergraph which is assumed to be a connected graph. However, such assumption cannot be ensured in some real-world networks. In particular, the hypergraph may become fragmented. That is, it may consist of a large number of connected components and isolated nodes, despite the fact that the original network is a connected graph.
Therefore, the existing higher-order methods would suffer seriously from the above fragmentation issue, since in these approaches, nodes without connection in hypergraph can't be grouped together even if they belong to the same community. To address the above fragmentation issue, we propose an \underline{Ed}ge enhancement approach for \underline{Mot}if-aware community detection (\textbf{EdMot}). The main idea is as follows. Firstly, a motif-based hypergraph is constructed and the top $K$ largest connected components in the hypergraph are partitioned into modules. Afterwards, the connectivity structure within each module is strengthened by constructing an edge set to derive a clique from each module. Based on the new edge set, the original connectivity structure of the input network is enhanced to generate a rewired network, whereby the motif-based higher-order structure is leveraged and the hypergraph fragmentation issue is well addressed. Finally, the rewired network is partitioned to obtain the higher-order community structure.
Extensive experiments have been conducted on eight real-world datasets and the results show the effectiveness of the proposed method in improving the community detection performance of state-of-the-art methods.
\end{abstract}

%
%



\maketitle

\section{Introduction}
\label{sec:intro}

Recent years have witnessed a growing trend in many different disciplines that model and interpret structured data as networks~\cite{Aggarwal2010Managing,DBLP:journals/pr/LiWLL18}. As ubiquitous abstractions depicting relationships among entities, networks have become a focus of data science and network analysis has attracted an increasing amount of attention from different fields such as physics, biology, mathematics and computer science~\cite{newman2004detecting,newman2006modularity,DBLP:conf/bibm/HuangWC18}. Community detection is an important task in network analysis that aims to partition the network into communities of strongly connected nodes.

Although many community detection methods have been proposed~\cite{shao2015community,he2016joint,Vincent2008Fast}, most of them only take into consideration the lower-order structure of the network at the level of individual nodes and edges, which fails to unravel the higher-order organization of the network.
Recently, to go beyond the lower-order connectivity patterns, some motif-based higher-order community detection methods have been proposed~\cite{tsourakakis2017scalable,benson2016higher,Li_HigherorderBrain:18,DBLP:conf/icdm/HuangWC18,li2018community,Huang_AAAI2019}, which can capture the motif-based characteristics and gain new insights into the organization of the network. However,
they typically focus on only the higher-order connections that are encoded in the motif-based hypergraph but underestimate and even violate the original lower-order topological structure. In particular, the hypergraph is assumed to be a connected graph in which the consequent partitioning procedure can be applied. However, in some real-world networks, such assumption cannot be ensured. That is, the hypergraph may be fragmented to a large number of connected components (with various sizes) and isolated nodes.
This is because the higher-order connections among nodes are built upon motifs. Two nodes are said to have higher-order connection if they have involved in at least one common motif and vise versa. It's worth noting that two nodes can be separated in the hypergraph of higher-order connections despite that they are connected in the original network. In this way, the number of connected components would increase and more isolated nodes would appear. These isolated nodes (from the perspective of higher-order connections) will render the community structure with instability. Therefore, the existing higher-order methods would suffer seriously from the above fragmentation issue since nodes without higher-order connections can't be grouped together even if they belong to the same community.

To address the above fragmentation issue, we propose an \underline{Ed}ge enhancement approach for \underline{Mot}if-aware community detection (\textbf{EdMot}). The main idea is as follows. Firstly, a motif-based hypergraph is constructed and the top $K$ largest connected components (measured by the number of contained nodes) in the hypergraph are partitioned into modules. Afterwards, the connectivity structure within each module is strengthened by constructing an edge set to derive a clique from each module. Based on the new edge set, the original connectivity structure of the input network is enhanced to generate a rewired network, whereby both the higher-order structure and lower-order structure are integrated. Finally, by partitioning the rewired network, the higher-order community structure can be discovered. In this study, we focus on the triangle motif for its ubiquitousness in social networks~\cite{holland1977method,newman2003social,prat2016put}, but our technique can be extended to other motifs as well.
Extensive experiments have been conducted on eight real-world datasets and the results show the effectiveness of the proposed method in improving the community detection performance of state-of-the-art methods.

We summarize the main contributions as follows:
\begin{enumerate}[{1)}]
\item
We present and formalize the hypergraph fragmentation issue suffered by the existing motif-based community detection methods, where higher-order connections are preserved but the original lower-order structure may be underestimated and even violated.
\item
We propose an \underline{Ed}ge enhancement approach for \underline{Mot}if-aware community detection (\textbf{EdMot}), which can not only leverage higher-order connections of the network but also overcome the hypergraph fragmentation issue.
\item
Extensive experiments are conducted on eight real-world datasets to show the effectiveness of the proposed method.
\end{enumerate}


\section{Background and Problem Statement}
\label{sec:background}


\begin{figure*}[!t]
\vskip-0.1in
\begin{center}
\centerline{ {\subfigure[The original network structure of the Cora dataset]
{\includegraphics[width=0.40\linewidth]{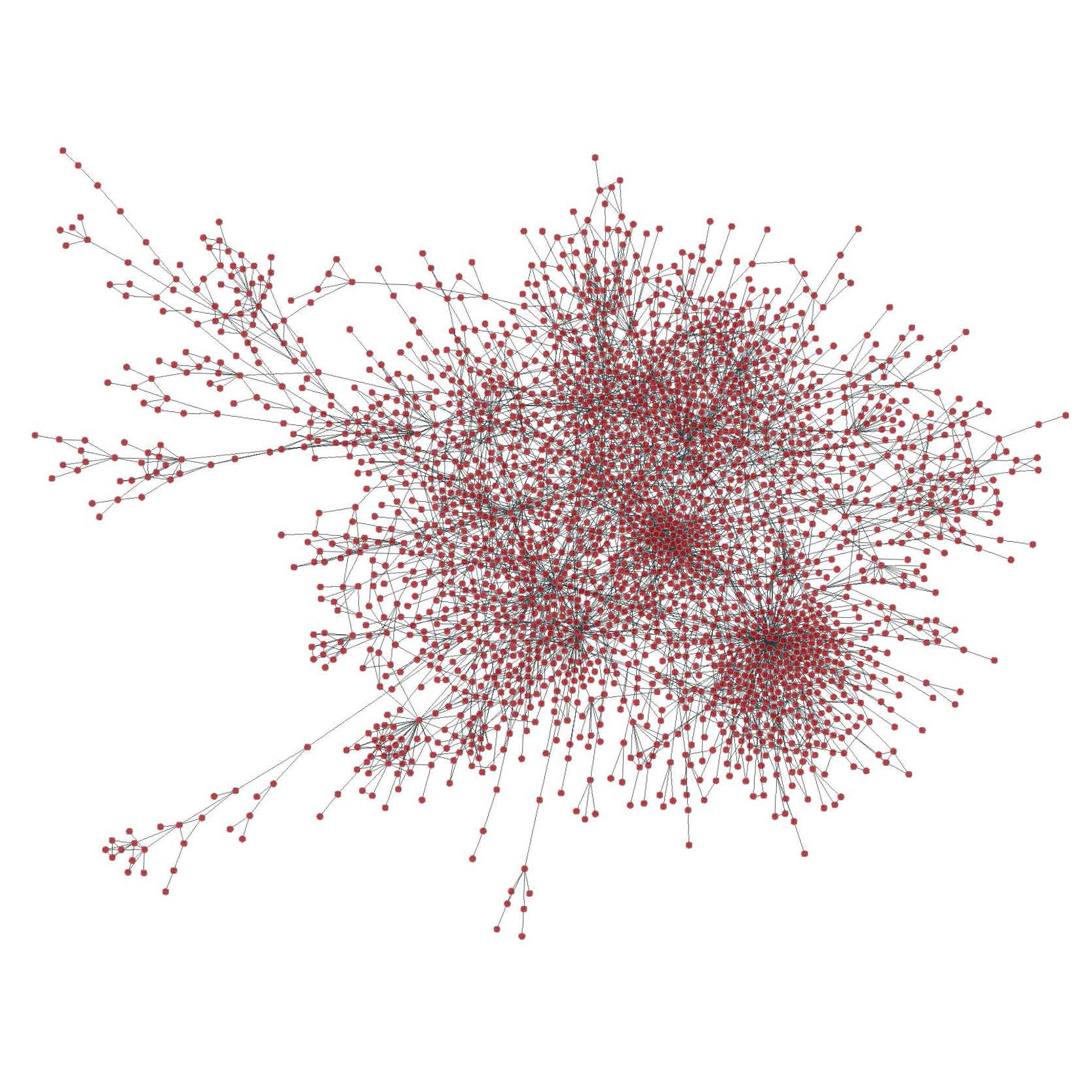}\label{fig:cora}}} \hskip -0.1in
{
{\includegraphics[width=0.15\linewidth]{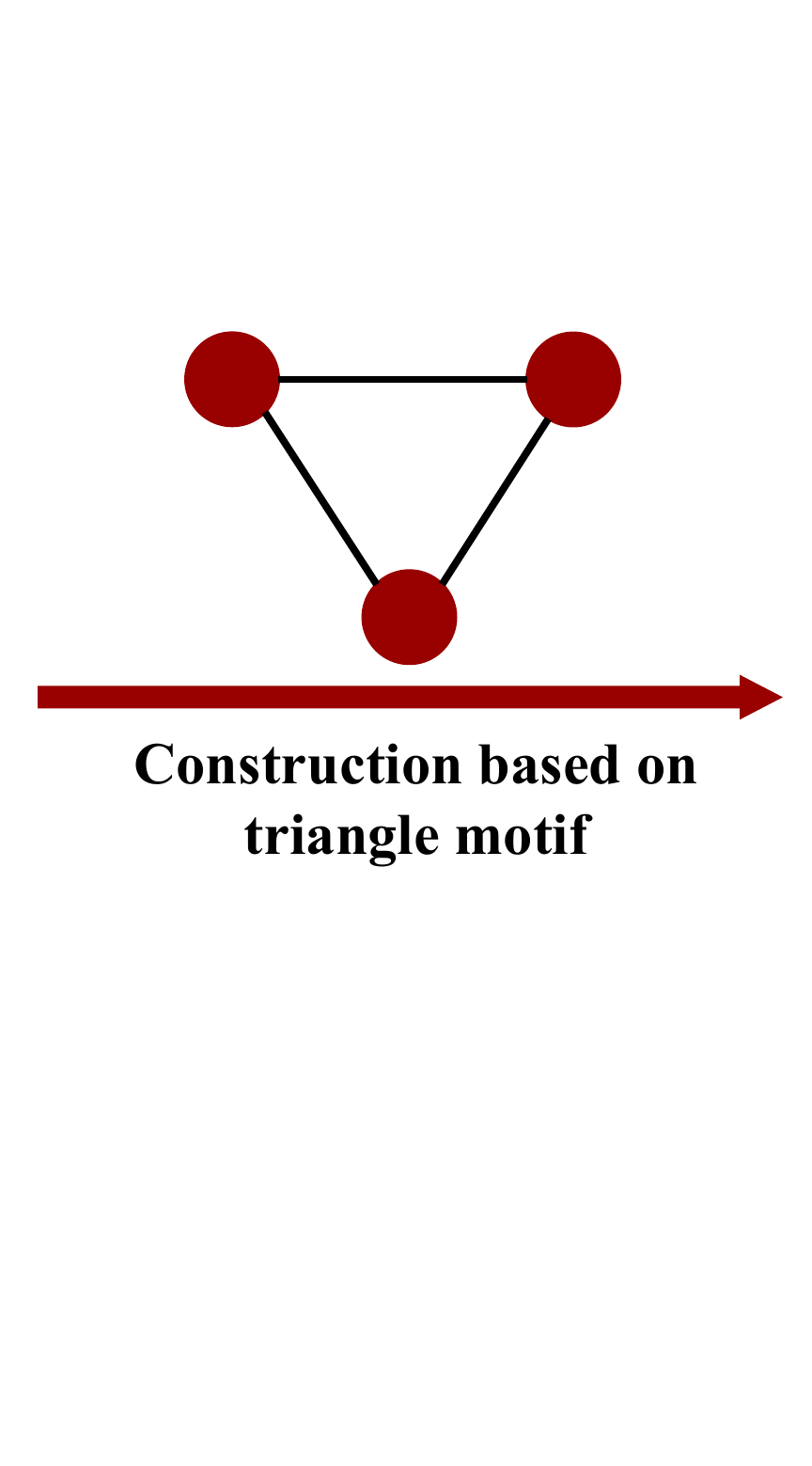}\label{fig:middle}}}\hskip -0.05in
{\subfigure[Motif-based hypergraph of the Cora dataset]
{\includegraphics[width=0.35\linewidth]{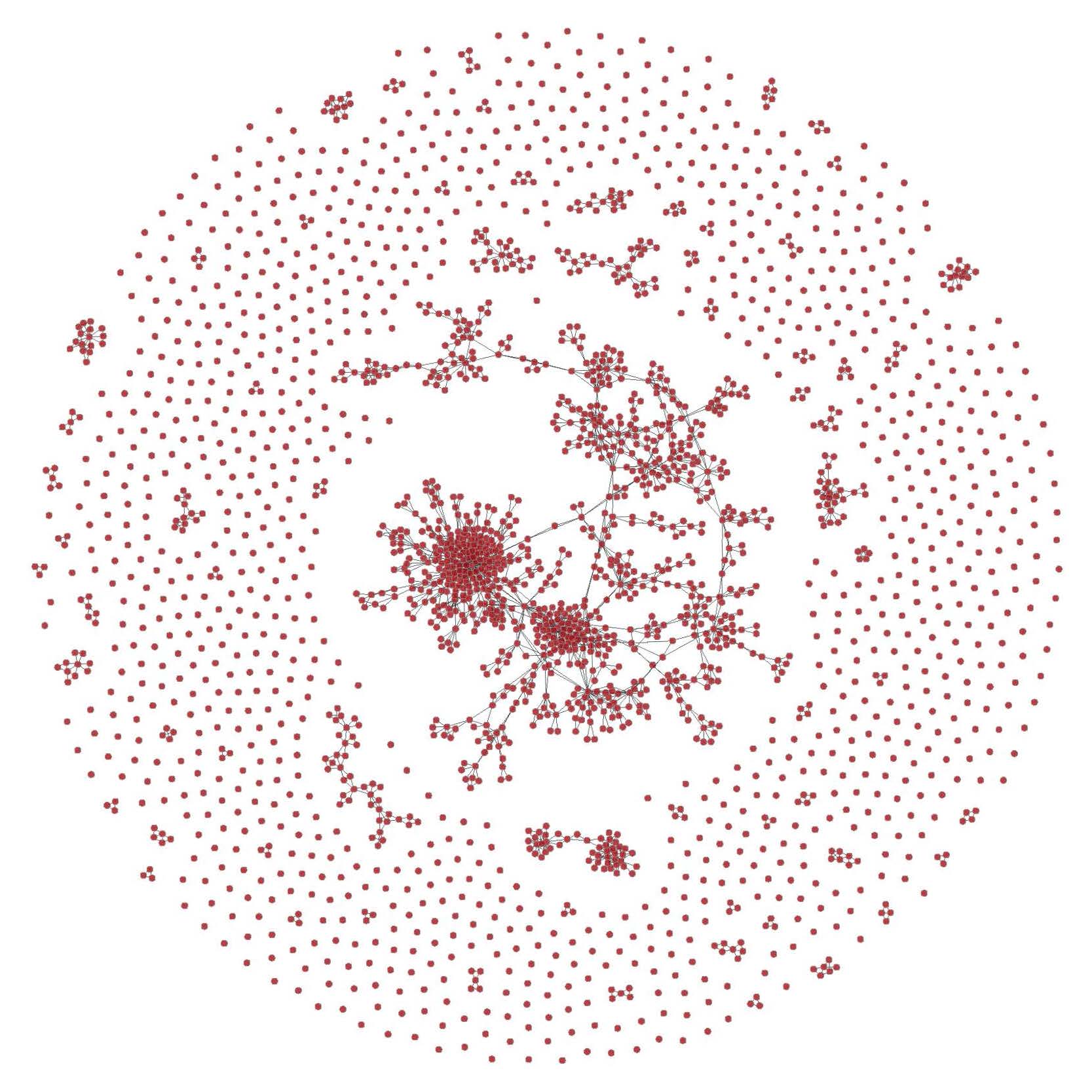}\label{fig:corawm4}}}
}
\end{center}
\vskip-0.1in
\caption{Illustration of the hypergraph fragmentation issue on the Cora dataset: The motif-based hypergraph constructed from the original network consists of several connected components (with various sizes) and a large number of isolated nodes.}
\label{fig:coracompare}
\vskip-0.1in
\end{figure*}

Assume that we are given a network $\mathcal{G}=\{\mathcal{V},\mathcal{E}\}$, where $\mathcal{V} = \{v_{i}|i=1,\dots,n\}$ represents the node set consisting of $n$ nodes and $\mathcal{E}=\{e_{i}|i=1,\dots,m\}$ represents the edge set consisting of $m$ undirected and unweighted edges. A node adjacency matrix $A\in \mathbb{R}^{n \times n}$ is used to encode the node-wise connection of the network, which is also known as the lower-order structure of the network~\cite{benson2016higher}. In this paper, the largest connected component of the original network will be extracted and utilized if it contains isolated nodes. We aim to infer the corresponding higher-order structure, which may characterize the building block of the network. In particular, the typical higher-order connectivity patterns, i.e., motifs, are identified to gain new insights into the organization of the network.
Formally, a network motif with $p$ nodes and $q$ edges can be denoted as:
\begin{equation}
\mathbf{M}_{p}^{q} = \{\mathcal{V}_\mathbf{M},\mathcal{E}_\mathbf{M}\}
\label{eq:motif}
\end{equation}
where ${\mathcal{V}_\mathbf{M}}\subseteq{\mathcal{V}}$ represents the set of $p$ nodes and ${\mathcal{E}_\mathbf{M}}\subseteq{\mathcal{E}}$ represents the set of $q$ edges. Following the convention of the literature~\cite{holland1977method,newman2003social,prat2016put}, we focus on $\mathbf{M}_{3}^{3}$, i.e., the triangle motif (as shown in the middle part of \figurename~\ref{fig:coracompare}). However, our technique can be well extended to other motifs. Conceptually, given the original node adjacency matrix $A$, the motif adjacency matrix constructed from the triangle motif, denoted as $W_{\mathbf{M}} \in \mathbb{R}^{n \times n}$, can be defined:
\begin{align}
\label{eq:Wm}
\mathit{(W_{\mathbf{M}})_{ij}} = \text{number of motif instances containing nodes $i$ and $j$.}
\end{align}
Note that $W_{\mathbf{M}}$ encodes the motif-based higher-order connections of the network, where edges correspond to co-occurrences in motifs and $(W_{\mathbf{M}})_{ij}$ can be 0 even though $A_{ij}>0$.

Several methods have been developed in terms of motif-based community detection~\cite{benson2016higher,tsourakakis2017scalable}. The common denominator of these methods is the dedication to leverage higher-order network structures effectively for community detection. However, they rely on the construction of a hypergraph whose edges correspond to motifs~\cite{benson2016higher}. In this way, the motif-based higher-order structure of the network is highlighted and well preserved but the original lower-order structure is underestimated and even violated. That is, edges that do not involve in any motifs would be eliminated. This leads to the serious issue called hypergraph fragmentation.

By ``fragmentation'', it means that the hypergraph or the motif adjacency matrix constructed from the original single connected component is usually fragmented into a large number of connected components with various sizes and isolated nodes due to the lack of higher-order connection, i.e., lack of involvement in the motif~\cite{zhao2015gsparsify}. As shown in~\figurename~\ref{fig:coracompare}, by constructing the motif-based hypergraph from the original network structure of the Cora dataset, a large number of connected components with various sizes and isolated nodes having no connection to any other nodes will be generated from the original single large connected component. As a consequence, most of the existing higher-order community detection methods that utilize the hypergraph would suffer from the hypergraph fragmentation issue~\cite{tsourakakis2017scalable,zhao2015gsparsify,benson2016higher}. In particular, by definition of communities (nodes should be densely connected within communities but sparsely connected between communities), the existing methods assume that the intermediate hypergraph is a single connected component~\cite{benson2016higher,tsourakakis2017scalable}, from which the eventual higher-order community structure can be discovered. However, as discussed above, the hypergraph is usually fragmented with a large number of connected components and isolated nodes. Therefore, the performance of the existing higher-order community detection methods can not be guaranteed.
For example, there will be a large number of dead nodes in the random walk process. Such isolated nodes would present confusing or rather, generate uncertain community memberships if the random label assignment strategy or the exclusion strategy is applied~\cite{kaiser2008mean}, leading to the degenerate performance.

A naive solution is directly grouping the connected components or isolated nodes together in the hypergraph to form the communities. Unfortunately, these components may be of different sizes and nodes with different ground-truth community labels may reside in the same component. For example, in~\figurename~\ref{fig:corawm4}, the largest connected component in the hypergraph consists of nodes from 7 different ground-truth communities. Furthermore, serious randomness would be introduced by the isolated nodes due to the lack of higher-order connections because there is no such community definition customized for isolated nodes. In this case, community memberships will be rendered with randomness if the isolated nodes are grouped randomly.

To address the above hypergraph fragmentation issue, we propose a novel method termed \textbf{EdMot} for motif-aware community detection. On the one hand, the motif-based higher-order structure is leveraged. On the other hand, the hypergraph fragmentation issue suffered by the traditional higher-order community detection methods can be well addressed by the newly designed edge enhancement strategy.


\section{The Proposed Method}


\subsection{Connected Component Identification}
We commence by constructing a motif adjacency matrix, (a.k.a. hypergraph) $W_{\mathbf{M}}$ through the original network structure to encode the higher-order connections. It can also be represented in the set form as follows:
\begin{equation}
\label{eq:GMotif}
\mathcal{G}^{\mathbf{M}} = \{\mathcal{V},\mathcal{E}^{\mathbf{M}} \}
\end{equation}
In the above equation, $\mathcal{G}^{\mathbf{M}}$ represents the motif-based hypergraph, $\mathcal{V}$ represents the node set that is the same as the original network, and $\mathcal{E}^{\mathbf{M}}$ represents the edge set consisting of $m_{w}$ weighted edges generated based on triangle motifs:
\begin{equation}
\label{eq:motifEdge}
\mathcal{E}^{\mathbf{M}} = \{(a,b,\tau)_{i}\}
\end{equation}
where $a,b\in \mathcal{V}$ are two end nodes of the $i$-th edge ($i \in \{1,\cdots, m_{w}\}$) and $\tau$ is the edge weight, i.e., the number of motif instances that contain node $a$ and node $b$ together. Accordingly, a set of $c_{\Phi}$ connected components of the hypergraph can be identified:
\begin{equation}
\label{eq:concomponents}
\Phi = \{ \phi_{i}\}
\end{equation}
where $\phi_{i}$ is the $i$-th connected component. That is, $\phi_{i} = \{\mathcal{V}^{\phi}_{i}, \mathcal{E}^{\mathbf{M}}_{i} \}$, $i \in \{1, \cdots,c_{\Phi}\}$, where $\mathcal{V}^{\phi}_{i}\subseteq \mathcal{V}$ is the set of nodes involved in the $i$-th connected component and $\mathcal{E}^{\mathbf{M}}_{i}\subseteq\mathcal{E}^{\mathbf{M}}$ is the weighted edge set of the $i$-th connected component respectively.

Apart from $c_{\Phi}$ connected components, there is a set of isolated nodes in the hypergraph, denoted as
\begin{equation}
\label{eq:Viso}
\mathcal{V}_{iso} = \mathcal{V} - \bigcup_{i \in \{1, \cdots, c_{\Phi}\}} \mathcal{V}^{\phi}_{i}
\end{equation}
Actually, motif structures are well preserved in these components, i.e., triangle motifs can certainly be found in the components containing 3 or more nodes.
The top $K$ largest connected components (measured by the number of contained nodes), i.e.,
$ \Phi_{K} \subseteq \Phi$ ($K<c_{\Phi}$) will be obtained. And each connected component in $\Phi_{K}$ will be further partitioned into modules\footnote{To avoid confusing the partitioning results in each connected component with the eventual partitioning results, the ``module'' is used to name the partitioning results in each connected component.} by using some traditional graph partitioning methods.
The influence of the parameter $K$ will be analyzed in later section.

\subsection{Connected Component Partitioning}
As an example, we adopt the Louvain~\cite{Vincent2008Fast} method that heuristically maximizes the well-known community structure evaluation measure called modularity~\cite{newman2006modularity,newman2004finding} to partition each connected component $\phi_{l}\in\Phi_{K}$ into modules. In particular, by taking each connected component $\phi_{l}$ as the input network, the modularity $Q$ can be given as~\cite{newman2006modularity}:
\begin{equation}
\label{eq:modularity}
Q = \frac{1}{4\mu}\sum_{ij}(A_{ij}-\frac{k_{i}k_{j}}{2\mu})(s_{i}s_{j}+1) = \frac{1}{4\mu}\sum_{ij}(A_{ij}-\frac{k_{i}k_{j}}{2\mu})s_{i}s_{j}
\end{equation}
where $k_{i}$ and $k_{j}$ are the degrees of nodes $i$ and $j$ respectively and $\mu = \frac{1}{2}\sum_{i}k_{i}$ is the total number of edges in the network. $s_{i}$ is the community label of node $i$. $\frac{k_{i}k_{j}}{2\mu}$ is the expected number of edges between node $i$ and node $j$ in the randomly rewired graph preserving the same degree distribution. $s_{i}s_{j}$ is equal to 1 if node $i$ and node $j$ belong to the same community and -1 otherwise. 


The output of the above modularity-based community detection procedure is the partitions (modules) in the $l$-th connected component $\phi_{l}\in\Phi_{K}$. By putting all the partitions (modules) of all the top $K$ largest connected components together, we can obtain a module set, denoted as $\{\mathcal{M}_{1},\cdots, \mathcal{M}_{\bar{m}}\}$, where $\bar{m}$ is the number of modules obtained by partitioning all the top $K$ largest connected components.
It is worthy noting that many other graph partitioning schemes can also be applied. 

\subsection{Network Rewiring via Edge Enhancement}


The module set $\{\mathcal{M}_{1},\cdots, \mathcal{M}_{\bar{m}}\}$ obtained in the previous subsection partially encodes the higher-order community structure. The reason for using ``partially'' is as follows.
\begin{enumerate}[{1)}]
\item \textbf{Property I}: If two nodes belong to the same module, it is likely that they have the same ground-truth community label as shown by the existing higher-order community detection approaches~\cite{arenas2008motif,benson2016higher}, i.e., discovering higher-order communities from each connected component in the hypergraph. However, it is also possible that two different modules may belong to the same ground-truth community.
\item \textbf{Property II}: The module set $\{\mathcal{M}_{1},\cdots, \mathcal{M}_{\bar{m}}\}$ obtained so far involves only part of nodes in the network since only the top $K$ largest connected components in the hypergraph are processed in the connected component partitioning.
\end{enumerate}

To completely reveal the higher-order community structure, the lower-order structure should be taken into account as a complement to the higher-order connections. To this end, an edge enhancement approach is developed to rewire the connectivity structure of the original network, whereby both the connectivity structure within each module in $\{\mathcal{M}_{1},\cdots, \mathcal{M}_{\bar{m}}\}$ is strengthened and the original lower-order structure is considered. In this way, not only the higher-order connections of the network can be leveraged but also the hypergraph fragmentation issue can be well addressed.

First of all, the connectivity structure within each module $\{\mathcal{M}_{1},\cdots, \mathcal{M}_{\bar{m}}\}$ is strengthened as follows. For the nodes that have already been partitioned into the same module, e.g., $\mathcal{M}_{i} \in \{\mathcal{M}_{1},\cdots, \mathcal{M}_{\bar{m}}\} (i \in \{1,\cdots, \bar{m}\})$, their connectivity is strengthened in line with the assumption that motif-based connection should have higher priority compared with the lower-order connection. This is because motif-based connection reflects the social transitivity and may encode more impressive characteristics of the network~\cite{wasserman1994social,strogatz2001exploring}. To this end, a clique is obtained for each module $\mathcal{M}_{i}\in\{\mathcal{M}_{1},\cdots, \mathcal{M}_{\bar{m}}\}$ by constructing an edge to each node pair in $\mathcal{M}_{i}$. In this way, a new set of edges is constructed, denoted as $\mathcal{E}^{*}_{mod}$,
\begin{equation}
\label{eq:ModEdgeSet}
\mathcal{E}^{*}_{mod}=\{(a,b)|\forall a,b\in\mathcal{M}_{i}, \forall i=1,\cdots,\bar{m}\}
\end{equation}
Notice that, the nodes within each module $\mathcal{M}_{i}$ are interconnected with each other by the strongest connectivity pattern, i.e. a clique structure, which is almost impossible to be destroyed in the consequent partitioning procedure. According to \textbf{Property I}, it is rational to establish such strong connectivity. However, according to \textbf{Property II}, it is also necessary to take into account the nodes residing out of the module set as well as the original lower-order connectivity pattern to overcome the fragmentation issue.

Therefore, based on the new edge set $\mathcal{E}^{*}_{mod}$, the original connectivity structure of the input network is enhanced to generate a rewired network
\begin{equation}
\label{eq:GMA}
\mathcal{G}_{A}^{\mathbf{M}} = \{\mathcal{V},\mathcal{E}_{A}^{\mathbf{M}}\} \text{ with~~} \mathcal{E}_{A}^{\mathbf{M}} = \mathcal{E} \cup \mathcal{E}^{*}_{mod}
\end{equation}
In this way, the edge enhanced network contains the same node set as the original network but encodes the connectivity patterns from both the original lower-order network structure in terms of $\mathcal{E}$ and the higher-order connections in terms of $\mathcal{E}^{*}_{mod}$.

\subsection{Method Summary and Computational Complexity}
\label{sec:complexity}

The rewired network, i.e., $\mathcal{G}_{A}^{\mathbf{M}}$, is fed into some graph partitioning methods to obtain the final community structure, i.e., $\mathcal{\widehat{C}} = \{ \mathcal{C}_{1}, \cdots, \mathcal{C}_{c}\}$, where $c$ is the number of communities in the final partitioning.

\begin{figure*}[!t]
\centering 
\includegraphics[width=0.85\linewidth]{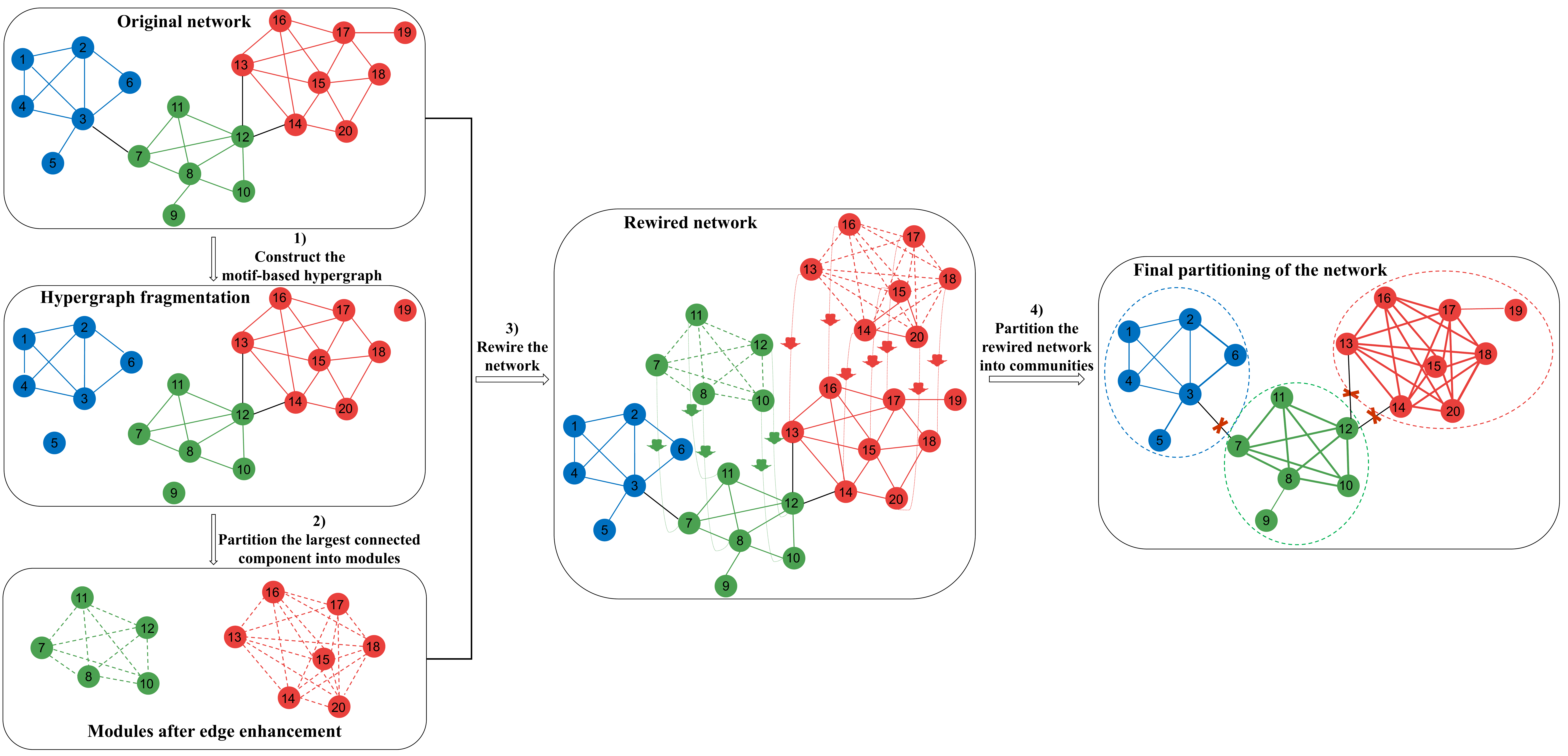}
\caption{Illustration of the proposed \textbf{EdMot} algorithm. A synthetic network is designed to serve as the original network, where nodes and edges in three communities are denoted with different colors and the black edges represent the inter-community edges. Specifically, by constructing the motif-based hypergraph in step 1, the hypergraph fragmentation issue arises, where two connected components and three isolated nodes are generated in the hypergraph. By partitioning the largest connected component into modules in step 2, two modules can be identified and a new edge set is constructed to derive a clique from each module, as shown as the dashed line. By rewiring the network in step 3, a rewired network can be obtained by substituting the new edge set into the original network. Finally, by partitioning the rewired network into communities, the community structure can be discovered.} \label{fig:diagall}
\end{figure*}

For clarification, Algorithm 1 summarizes the main procedure of the proposed \textbf{EdMot} method. In addition,~\figurename~\ref{fig:diagall} exhibits the whole process intuitively.

\begin{algorithm}[htb]
\caption{The proposed \textbf{EdMot} approach.}
\label{alg:Framwork}
{\bfseries Input:} Original network $A\in \mathbb{R}^{n \times n}$, parameter $K$, a graph partitioning method \texttt{S}.
\begin{algorithmic}[1]
\STATE Construct motif adjacency matrix $W_{\mathbf{M}}$ from $A$ via (\ref{eq:Wm}).\\
\STATE Obtain connected components $\Phi$ via (\ref{eq:concomponents}) and select top $K$ largest connected components $\Phi_{K} \subseteq \Phi $.\\
\STATE Apply \texttt{S} to partition each $\phi_{l} \in \Phi_{K}$, $\forall l \in \{1,\cdots, K\}$ and aggregate the results to form the module set $\{\mathcal{M}_{1},\cdots, \mathcal{M}_{\bar{m}}\}$.\\
\STATE Construct a new edge set $\mathcal{E}^{*}_{mod}$ via (\ref{eq:ModEdgeSet}).\\
\STATE Rewire the original network to obtain $\mathcal{G}_{A}^{\mathbf{M}}$ via (\ref{eq:GMA}).\\
\STATE Feed $\mathcal{G}_{A}^{\mathbf{M}}$ into \texttt{S} to obtain final community structure $\mathcal{\widehat{C}}$.\\
\end{algorithmic}
\leftline{{\bfseries Output:} Community structure $\mathcal{\widehat{C}}$.}
\end{algorithm}

We now analyze the computational complexity of the proposed \textbf{EdMot} method. Overall, the complexity of the method is governed by the computations of the motif adjacency matrix $W_{\mathbf{M}}$ and the graph partitioning in both the connected components and rewired network. For simplicity, we assume that we can access the edges in a graph in $O(1)$ time and can access and modify the elements in the matrix in $O(1)$ time. The computational time of constructing $W_{\mathbf{M}}$ is bounded by the time to find all the motif instances in the graph. Theoretically, we can compute $W_{\mathbf{M}}$ in $O(n^{p})$ time for a motif with $p$ nodes. However, most real-world networks are sparse, we can instead focus on the computational complexity in terms of the number of edges in the network. In particular, for the triangle motifs discussed in this paper, the motif instances can be found in $O(m^{1.5})$ time~\cite{latapy2008main,Berry2014Why,benson2016higher}. As for the graph partitioning, in this paper, we adopt the heuristic modularity maximization method as an example, which can be finished in $O(n\log n)$ on average for its ability to find the hierarchical community structure~\cite{Vincent2008Fast}. Therefore, the overall computational complexity is $O(m^{1.5}+n\log n)$.

\section{Experiments}
\label{sec:experiments}
In this section, extensive experiments are conducted to confirm the effectiveness of the proposed method. The Matlab code is available in https://github.com/lipzh5/EdMot\_pro.git.

\subsection{Datasets and Evaluation Measures}

In our experiments, eight real-world datasets are used (the edges are treated to be undirected for all the datasets). On the first four datasets, the ground-truth community labels are provided for evaluation purpose and on the rest four datasets, there is no ground-truth community information where the internal evaluation measures such as modularity are used for evaluation. 

Datasets with ground-truth community labels:
\begin{enumerate}[{1)}]
\item
\textbf{polbooks\footnotemark[1].} A network of books about US politics, where edges between books represent frequent copurchasing of books by the same buyers. It consists of 105 nodes and 441 edges.
\item
\textbf{email-Eu-core\footnotemark[2].} A email network of communication between institution members, which is generated using email data from a large European research institution. It consists of 1005 nodes and 25571 edges.
\item
\textbf{polblogs\footnotemark[1].} A network of hyperlinks between weblogs on US politics, which is recorded in 2005. It consists of 1490 nodes and 19090 edges.
\item
\textbf{Cora\footnotemark[3].} A citation network of scientific publications, which consists of maching learning papers that can be classified into seven classes. It consists of 2708 nodes and 5429 edges.
\end{enumerate}

Datasets without ground-truth community labels:
\begin{enumerate}[{1)}]
\item
\textbf{power\footnotemark[4].} A network that represents the topology of the Western States Power Grid of the United States. It consists of 4941 nodes and 6594 edges.
\item
\textbf{ca-GrQc\footnotemark[2].} A collaboration network of scientific collaborations between authors with papers submitted to General Relativity and Quantum Cosmology category. It consists of 5242 nodes and 14496 edges.
\item
\textbf{as-22july06\footnotemark[4].} A symmetric snapshot of the structure of the Internet at the level of autonomous systems. It consists of 22963 nodes and 48436 edges.
\item
\textbf{email-Enron\footnotemark[4].} A email network that covers all the email communication within a dataset of around half million emails. It consists of 36692 nodes and 367662 edges.
\end{enumerate}
\footnotetext[1]{\url{http://www-personal.umich.edu/~mejn/netdata/}}
\footnotetext[2]{\url{http://snap.stanford.edu/data/}}
\footnotetext[3]{\url{http://linqs.cs.umd.edu/projects/projects/lbc/}}
\footnotetext[4]{\url{https://graph-tool.skewed.de/static/doc/collection.html}}

Three commonly used evaluation measures are adopted for evaluating the quality of the discovered community structure, namely Normalized Mutual Information (NMI), F-score and Modularity (Q)~\cite{chakraborty2014permanence}. The first two measures require the ground-truth community labels for evaluation purpose and the values are in the range between 0 and 1. The last measure $Q$ ranges from -1 to 1. For all the three measures, a higher value indicates better performance.

\subsection{Graph Partitioning Methods}
\label{sec:clusteringmethod}
We adopt the following four graph partitioning methods in our experiments:
\begin{enumerate}[1)]
\item
\textbf{Louvain}~\cite{Vincent2008Fast}: It is a greedy community detection method that can reveal the hierarchical community structure.
\item
\textbf{Spectral Clustering (SC)}~\cite{ng2002spectral}: It performs a spectral clustering of the node adjacency matrix into $c$ clusters.
\item
\textbf{Affinity Propagation (AP)}~\cite{frey2007clustering}: It is a fast clustering method based on similarities between pairs of data points.
\item
\textbf{Nonnegative Matrix Factorization (NMF)}~\cite{cai2011graph}: It obtains the new property representation by factorizing the node adjacency matrix into two nonnegative matrices.
\end{enumerate}
Each of the above four graph partitioning methods is taken as the input ``graph partitioning method \texttt{S}'' of Algorithm~\ref{alg:Framwork}, by which different variants of \textbf{EdMot} are obtained, namely EdMot-Louvain, EdMot-SC, EdMot-AP and EdMot-NMF. Similarly, as the conventional higher-order community detection methods, each of the above four graph partitioning methods takes as input the hypergraph of higher-order connectivity, i.e. the motif adjacency matrix $W_{\mathbf{M}}$, by which different variants of the higher-order community detection methods are obtained, namely Motif-Louvain, Motif-SC, Motif-AP and Motif-NMF. And the original graph partitioning method, the corresponding higher-order variant, and the corresponding \textbf{EdMot} variant are compared and analyzed. For instance, Louvain, Motif-Louvain and EdMot-Louvain are compared.

\subsection{Comparison Results}
\label{sec:results}

\begin{table*}[!t]
\caption{Comparison results on the Louvain method. The best result in each measure is highlighted in bold.}
\label{table:Louvaincompare}
\begin{center}
\vskip -0.1in
\begin{tabular}{c|c|c|c|c|c|c}
\hline
\multicolumn{2}{c|}{~~~} &polbooks    &email-Eu-core  &polblogs &Cora  &Avg.rank\\
\hline
\multirow{3}*{NMI}&Louvain &0.4142\scriptsize{$\pm0.0000$}   &0.5642\scriptsize{$\pm0.0000$}   &0.2684\scriptsize{$\pm0.0000$}   &0.3996\scriptsize{$\pm0.0000$} &2.75\\
\cline{2-6}
& Motif-Louvain &0.4818\scriptsize{$\pm0.0000$}   &$\mathbf{0.6350}$\scriptsize{$\pm0.0000$}   &0.2513\scriptsize{$\pm0.0000$}   &0.4043\scriptsize{$\pm0.0000$} &2.00\\
\cline{2-6}
& EdMot-Louvain &$\mathbf{0.4981}$\scriptsize{$\pm0.0000$}   &0.6098\scriptsize{$\pm0.0000$}   &$\mathbf{0.3464}$\scriptsize{$\pm0.0000$}   &$\mathbf{0.4088}$\scriptsize{$\pm0.0000$} &1.25\\
\hline
\multirow{3}*{F-score}& Louvain &0.4954\scriptsize{$\pm0.0000$}   &0.5430\scriptsize{$\pm0.0000$}   &0.6117\scriptsize{$\pm0.0000$}  &0.0486\scriptsize{$\pm0.0000$} &2.50 \\
\cline{2-6}
&Motif-Louvain &0.6340\scriptsize{$\pm0.0000$}   &0.5149\scriptsize{$\pm0.0000$}   &0.5858\scriptsize{$\pm0.0000$}   &0.0517\scriptsize{$\pm0.0000$} &2.50\\
\cline{2-6}
& EdMot-Louvain &$\mathbf{0.6462}$\scriptsize{$\pm0.0000$}   &$\mathbf{0.5735}$\scriptsize{$\pm0.0000$}   &$\mathbf{0.7613}$\scriptsize{$\pm0.0000$}   &$\mathbf{0.0690}$\scriptsize{$\pm0.0000$} &1.00\\
\hline
\multirow{3}*{Modularity}& Louvain &0.4833\scriptsize{$\pm0.0000$}   &0.3933\scriptsize{$\pm0.0000$}   &0.4272\scriptsize{$\pm0.0000$}   &0.5439\scriptsize{$\pm0.0000$} &2.50 \\
\cline{2-6}
&Motif-Louvain &0.5058\scriptsize{$\pm0.0000$}   &0.4024\scriptsize{$\pm0.0000$}   &0.4221\scriptsize{$\pm0.0000$}   &0.4216\scriptsize{$\pm0.0000$} &2.50 \\
\cline{2-6}
& EdMot-Louvain &$\mathbf{0.5092}$\scriptsize{$\pm0.0000$}   &$\mathbf{0.4085}$\scriptsize{$\pm0.0000$}   &$\mathbf{0.4307}$\scriptsize{$\pm0.0000$}   &$\mathbf{0.5783}$\scriptsize{$\pm0.0000$} &1.00\\
\hline
\end{tabular}
\end{center}
\end{table*}

\begin{table*}[!t]
\caption{Comparison results on the SC method. The best result in each measure is highlighted in bold.}
\label{table:SCcompare}
\begin{center}
\vskip -0.1in
\begin{tabular}{c|c|c|c|c|c|c}
\hline
\multicolumn{2}{c|}{~~~} &polbooks    &email-Eu-core  &polblogs &Cora  &Avg.rank  \\
\hline
\multirow{3}*{NMI}&SC &0.5422\scriptsize{$\pm0.0000$}   &$\mathbf{0.7049}$\scriptsize{$\pm0.0034$}   &0.0016\scriptsize{$\pm0.0000$}   &0.3953\scriptsize{$\pm0.0000$} &{2.25} \\
\cline{2-6}
& Motif-SC &0.5458\scriptsize{$\pm0.0102$}  &0.6509\scriptsize{$\pm0.0051$}   &0.0030\scriptsize{$\pm0.0016$}  &0.0913\scriptsize{$\pm0.0366$} &2.50\\
\cline{2-6}
& EdMot-SC &$\mathbf{0.5675}$\scriptsize{$\pm0.0000$}   &0.6913\scriptsize{$\pm0.0022$}   &$\mathbf{0.3975}$\scriptsize{$\pm0.0000$}   &$\mathbf{0.4226}$\scriptsize{$\pm0.0038$} &1.25\\
\hline
\multirow{3}*{F-score}& SC &0.8216\scriptsize{$\pm0.0000$}   &$\mathbf{0.5533}$\scriptsize{$\pm0.0076$}   &0.7580\scriptsize{$\pm0.0000$}   &0.5392\scriptsize{$\pm0.0000$} &{2.00}\\
\cline{2-6}
& Motif-SC &0.8298\scriptsize{$\pm0.0059$}   &0.4896\scriptsize{$\pm0.0076$}   &0.6541\scriptsize{$\pm0.0634$}   &0.3562\scriptsize{$\pm0.0185$} &{2.75}\\
\cline{2-6}
& EdMot-SC &$\mathbf{0.8416}$\scriptsize{$\pm0.0000$}   &0.5419\scriptsize{$\pm0.0027$}   &$\mathbf{0.8112}$\scriptsize{$\pm0.0000$}   &$\mathbf{0.5728}$\scriptsize{$\pm0.0265$} &1.25\\
\hline
\multirow{3}*{Modularity}& SC &0.5015\scriptsize{$\pm0.0000$}   &$\mathbf{0.2598}$\scriptsize{$\pm0.0029$}   &0.0007\scriptsize{$\pm0.0000$}   &0.6599\scriptsize{$\pm0.0000$} &2.25\\
\cline{2-6}
& Motif-SC &0.5041\scriptsize{$\pm0.0032$}   &0.2410\scriptsize{$\pm0.0033$}   &0.0009\scriptsize{$\pm0.0003$}  &0.3674\scriptsize{$\pm0.0569$}  &2.50\\
\cline{2-6}
& EdMot-SC &$\mathbf{0.5099}$\scriptsize{$\pm0.0000$}   &0.2520\scriptsize{$\pm0.0014$}   &$\mathbf{0.4309}$\scriptsize{$\pm0.0000$}   &$\mathbf{0.7108}$\scriptsize{$\pm0.0103$} &1.25\\
\hline
\end{tabular}
\end{center}
\end{table*}

\begin{table*}[!t]
\caption{Comparison results on the AP method. The best result in each measure is highlighted in bold.}
\label{table:APcompare}
\begin{center}
\vskip -0.1in
\begin{tabular}{c|c|c|c|c|c|c}
\hline
\multicolumn{2}{c|}{~~~} &polbooks    &email-Eu-core  &polblogs &Cora &Avg.rank\\
\hline
\multirow{3}*{NMI}&AP &0.3541\scriptsize{$\pm0.0138$}  &0.4519\scriptsize{$\pm0.0149$}   &0.1238\scriptsize{$\pm0.0024$}   &0.3856\scriptsize{$\pm0.0015$} &3.00 \\
\cline{2-6}
& Motif-AP &0.3974\scriptsize{$\pm0.0077$}  &$\mathbf{0.6318}$\scriptsize{$\pm0.0011$}   &$\mathbf{0.1772}$\scriptsize{$\pm0.0003$}   &0.3962\scriptsize{$\pm0.0010$} &1.50 \\
\cline{2-6}
& EdMot-AP &$\mathbf{0.4029}$\scriptsize{$\pm0.0044$}  &0.5909\scriptsize{$\pm0.0062$}   &0.1601\scriptsize{$\pm0.0028$}   &$\mathbf{0.3973}$\scriptsize{$\pm0.0021$} &1.50 \\
\hline
\multirow{3}*{F-score}& AP &0.2510\scriptsize{$\pm0.0116$}  &0.2075\scriptsize{$\pm0.0144$}   &0.0983\scriptsize{$\pm0.0026$}   &0.0666\scriptsize{$\pm0.0016$} &{2.50} \\
\cline{2-6}
& Motif-AP &0.2748\scriptsize{$\pm0.0211$}  &0.2219\scriptsize{$\pm0.0014$}   &$\mathbf{0.1046}$\scriptsize{$\pm0.0006$}   &0.0481\scriptsize{$\pm0.0009$} &2.00 \\
\cline{2-6}
& EdMot-AP &$\mathbf{0.2873}$\scriptsize{$\pm0.0129$}  &$\mathbf{0.3185}$\scriptsize{$\pm0.0053$}   &0.0980\scriptsize{$\pm0.0117$}   &$\mathbf{0.0750}$\scriptsize{$\pm0.0046$} &{1.50} \\
\hline
\multirow{3}*{Modularity}& AP &0.2214\scriptsize{$\pm0.0151$}  &0.0749\scriptsize{$\pm0.0065$}   &0.1045\scriptsize{$\pm0.0026$}   &0.4723\scriptsize{$\pm0.0026$}  &2.75 \\
\cline{2-6}
& Motif-AP &0.2977\scriptsize{$\pm0.0114$}  &0.1712\scriptsize{$\pm0.0010$}   &$\mathbf{0.2251}$\scriptsize{$\pm0.0009$} &0.3821\scriptsize{$\pm0.0024$} &{2.00} \\
\cline{2-6}
& EdMot-AP &$\mathbf{0.3069}$\scriptsize{$\pm0.0159$}  &$\mathbf{0.2187}$\scriptsize{$\pm0.0029$}   &0.1148\scriptsize{$\pm0.0098$}   &$\mathbf{0.4770}$\scriptsize{$\pm0.0321$} &{1.25} \\
\hline
\end{tabular}
\end{center}
\end{table*}

\begin{table*}[!t]
\caption{Comparison results on the NMF method. The best result in each measure is highlighted in bold.}
\label{table:NMFcompare}
\begin{center}
\vskip -0.1in
\begin{tabular}{c|c|c|c|c|c|c}
\hline
\multicolumn{2}{c|}{~~~} &polbooks    &email-Eu-core  &polblogs &Cora &Avg.rank\\
\hline
\multirow{3}*{NMI}&NMF &0.4528\scriptsize{$\pm0.0086$}   &0.7090\scriptsize{$\pm0.0063$}   &0.4005\scriptsize{$\pm0.0000$}   &0.2723\scriptsize{$\pm0.0071$}&2.25 \\
\cline{2-6}
& Motif-NMF &0.4809\scriptsize{$\pm0.0302$}   &0.6594\scriptsize{$\pm0.0060$}   &0.2687\scriptsize{$\pm0.0100$}   &0.1010\scriptsize{$\pm0.0058$} &2.75 \\
\cline{2-6}
& EdMot-NMF &$\mathbf{0.5568}$\scriptsize{$\pm0.0042$}   &$\mathbf{0.7113}$\scriptsize{$\pm0.0042$}   &$\mathbf{0.4040}$\scriptsize{$\pm0.0012$}   &$\mathbf{0.2821}$\scriptsize{$\pm0.0028$} &1.00 \\
\hline
\multirow{3}*{F-score}& NMF &0.7750\scriptsize{$\pm0.0101$}   &0.4756\scriptsize{$\pm0.0016$}   &0.7644\scriptsize{$\pm0.0000$}  &0.7644\scriptsize{$\pm0.0016$} &{2.00} \\
\cline{2-6}
& Motif-NMF &0.7326\scriptsize{$\pm0.0449$}   &$\mathbf{0.7644}$\scriptsize{$\pm0.0119$}   &0.7644\scriptsize{$\pm0.0050$}   &0.7644\scriptsize{$\pm0.0094$} &1.75 \\
\cline{2-6}
& EdMot-NMF &$\mathbf{0.8382}$\scriptsize{$\pm0.0024$}   &0.5811\scriptsize{$\pm0.0019$}   &$\mathbf{0.8145}$\scriptsize{$\pm0.0003$}   &0.4854\scriptsize{$\pm0.0033$} &{1.50} \\
\hline
\multirow{3}*{Modularity}& NMF &0.4484\scriptsize{$\pm0.0060$}   &0.2663\scriptsize{$\pm0.0055$}   &0.4305\scriptsize{$\pm0.0000$}   &$\mathbf{0.6499}$\scriptsize{$\pm0.0010$} &2.00 \\
\cline{2-6}
& Motif-NMF &0.4808\scriptsize{$\pm0.0078$}   &0.2729\scriptsize{$\pm0.0072$}   &0.4219\scriptsize{$\pm0.0005$}  &0.4407\scriptsize{$\pm0.0105$} &2.25 \\
\cline{2-6}
& EdMot-NMF &$\mathbf{0.4999}$\scriptsize{$\pm0.0019$}   &$\mathbf{0.2740}$\scriptsize{$\pm0.0087$}   &0.4305\scriptsize{$\pm0.0000$}   &0.6216\scriptsize{$\pm0.0099$} &1.25 \\
\hline
\end{tabular}
\end{center}
\end{table*}

%

Comparison results are reported from Table~\ref{table:Louvaincompare} to Table~\ref{table:modularity2}, where the scores are averaged over 20 runs for every method and the standard deviations are also reported. Table~\ref{table:Louvaincompare} to Table~\ref{table:NMFcompare} report the results on datasets with ground-truth community labels while Table~\ref{table:modularity2} is for two datasets without ground-truth community labels. Additionally, the average rank is also provided in the last column, which is computed by averaging the ranking positions of each method across the testing datasets.

Specifically, Table~\ref{table:Louvaincompare} reports the comparison results among Louvain, Motif-Louvain and EdMot-Louvain. As can be seen, the best scores are achieved by EdMot-Louvain in terms of NMI, F-score and modularity on polbooks, polblogs and Cora. On average, about 17\% and 14\% improvements (in terms of NMI) have been achieved by EdMot-Louvain over Louvain and Motif-Louvain respectively. This may be due to the utilization of edge enhancement and the original network structure, which plays an important role in avoiding hypergraph fragmentation and preserving the structural information of the network.
On the email-Eu-core dataset, the proposed method performs worse than Motif-Louvain in terms of NMI. The reason is that this dataset has a relatively denser linkage structure (i.e., 1005 nodes and 25571 edges) and the hypergraph fragmentation issue does not appear on this dataset. That is, the motif-based hypergraph is a single connected component. While the polblogs and Cora datasets suffer from the hypergraph fragmentation issue, which accounts for the better performance achieved by the proposed method. As for the polbooks dataset, even though the hypergraph fragmentation does not appear, the proposed method still performs well. Similar analysis can be made in Table~\ref{table:SCcompare} to Table~\ref{table:modularity2}.

In summary, the proposed method can perform better than the original graph partitioning methods and the traditional higher-order methods with hypergraph fragmentation issue. What's more, it may also improve the performance of some state-of-the-art graph partitioning methods despite there is no hypergraph fragmentation issue.

\begin{table}[!t]
\caption{Comparison results in terms of modularity on the two datasets without ground-truth community labels.}
\label{table:modularity2}
\begin{center}
\vskip -0.1in
\begin{tabular}{@{}c|c|c|c@{}}
  \hline
Method  &power &ca-GrQc&Avg.rank\\
\hline
  Louvain  &0.5248\scriptsize{$\pm0.0000$} & 0.6863\scriptsize{$\pm0.0000$} &3.00\\
  Motif-Louvain &0.8053\scriptsize{$\pm0.0000$} &0.7806\scriptsize{$\pm0.0000$} &2.00\\
  EdMot-Louvain  &$\mathbf{0.9983}$\scriptsize{$\pm0.0000$} & $\mathbf{0.9988}$\scriptsize{$\pm0.0007$} &1.00\\
  \hline
  SC & 0.8545\scriptsize{$\pm0.0000$}& 0.1426\scriptsize{$\pm0.0037$} &2.50\\
  Motif-SC &0.8159\scriptsize{$\pm0.0034$} &0.6725\scriptsize{$\pm0.0007$} &2.50\\
  EdMot-SC & $\mathbf{0.8607}$\scriptsize{$\pm0.0023$} & $\mathbf{0.7279}$\scriptsize{$\pm0.0027$}&1.00 \\
  \hline
  AP  & 0.4907\scriptsize{$\pm0.0020$}  &0.4220\scriptsize{$\pm0.0000$} &2.00\\
  Motif-AP & 0.1799\scriptsize{$\pm0.0005$}     &0.4169\scriptsize{$\pm0.0035$}&3.00\\
  EdMot-AP & $\mathbf{0.8551}$\scriptsize{$\pm0.0042$}    &$\mathbf{0.5483}$\scriptsize{$\pm0.0064$} &1.00\\
  \hline
  NMF  & 0.4922\scriptsize{$\pm0.0015$}    & 0.5433\scriptsize{$\pm0.0051$} &3.00\\
  Motif-NMF & 0.6336\scriptsize{$\pm0.0191$}    &0.5704\scriptsize{$\pm0.0171$} &2.00\\
  EdMot-NMF & $\mathbf{0.9406}$\scriptsize{$\pm0.0151$}    & $\mathbf{0.8920}$\scriptsize{$\pm0.0058$} &1.00\\
  \hline
\end{tabular}
\end{center}
\end{table}

\subsection{Parameter Analysis}
\label{sec:parameter}

\begin{figure}[!t]
\vskip -0.1in
\renewcommand{\subfigcapskip}{-4pt}
\renewcommand{\subfigbottomskip}{0pt}
\centerline{
{\subfigure[On the polblogs dataset.]
{\includegraphics[width=0.5\linewidth]{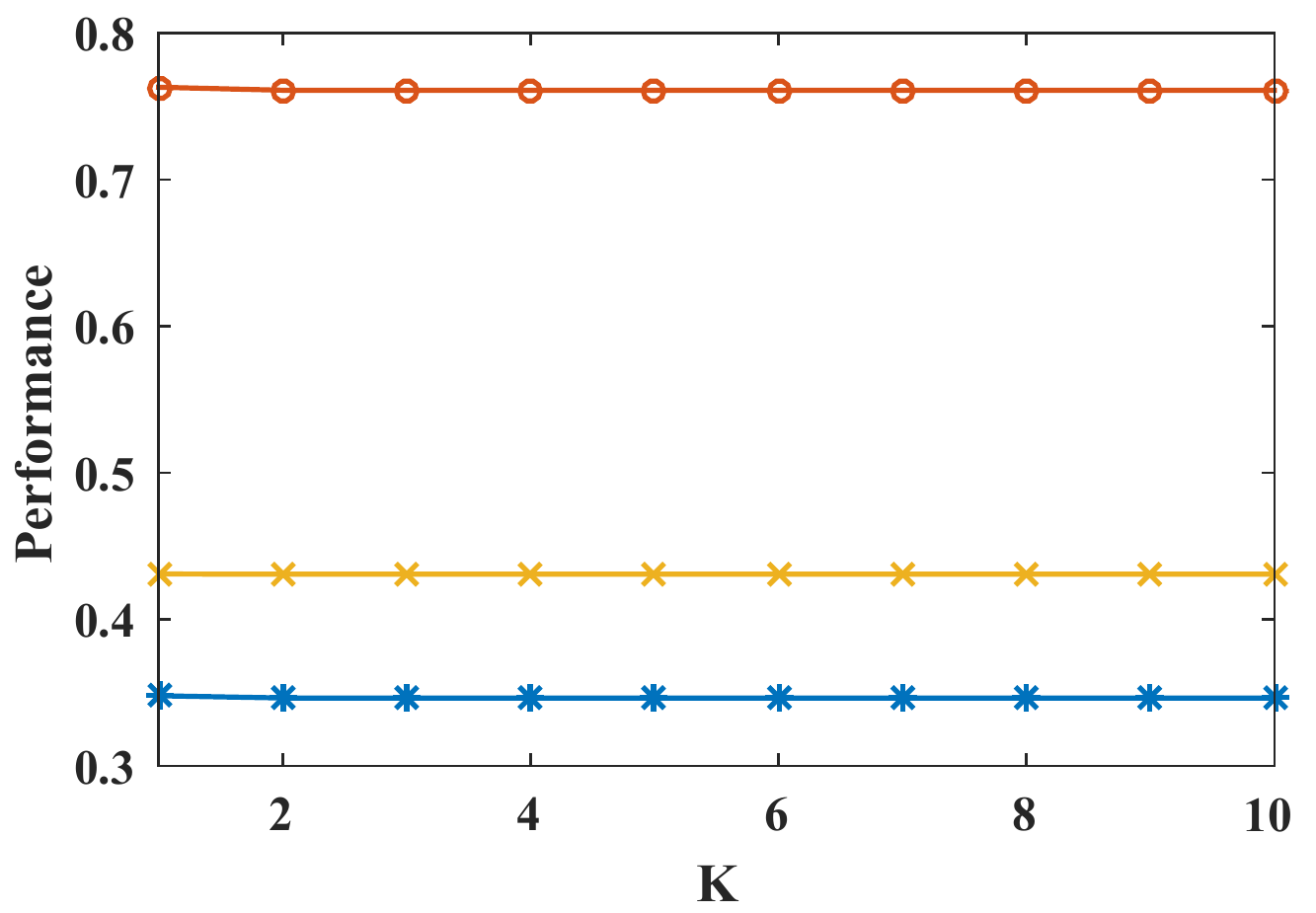}\label{fig:paraPolblog}}}
{\subfigure[On the Cora dataset.]
{\includegraphics[width=0.5\linewidth]{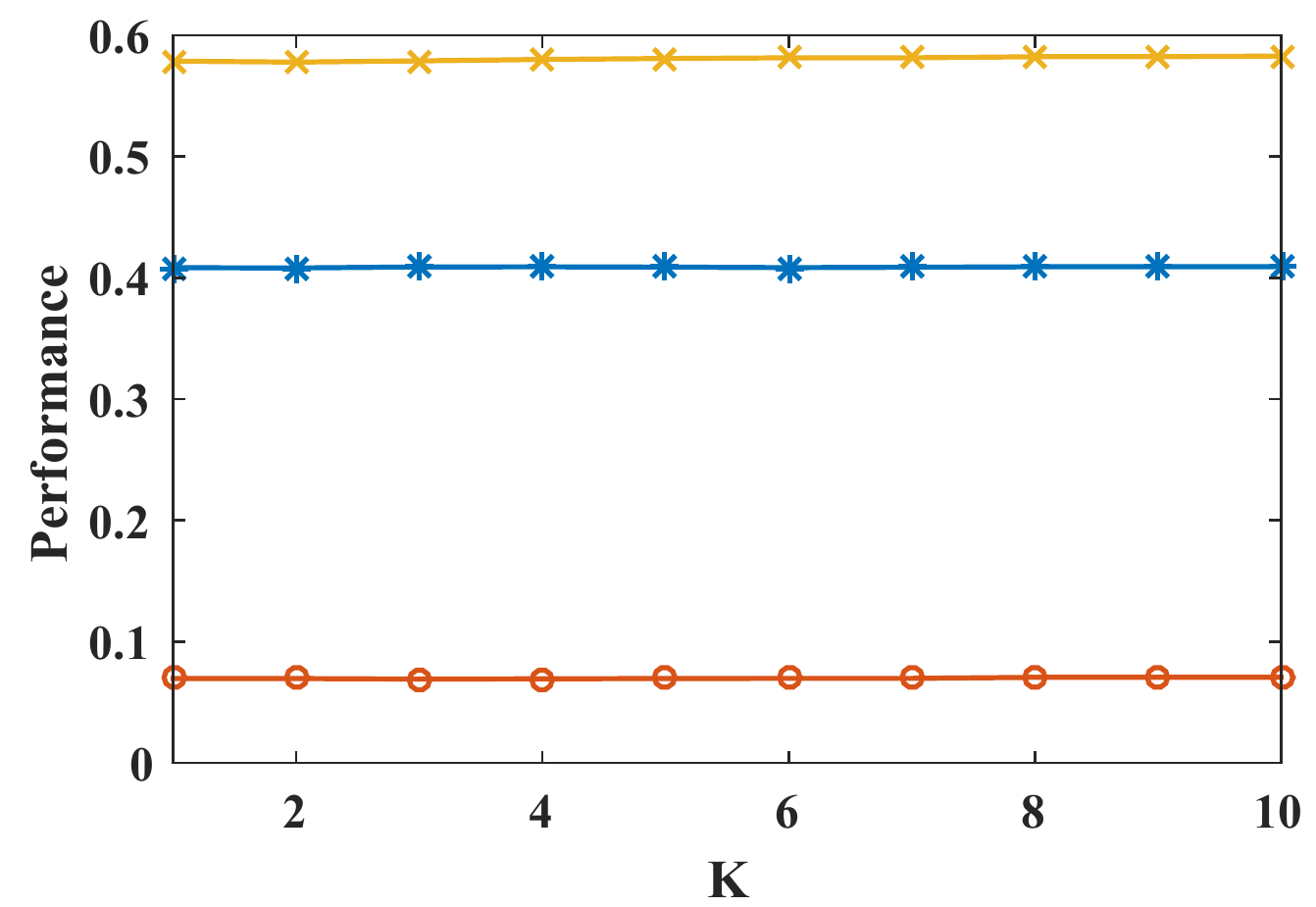}\label{fig:paraCora}}}}
\centering{\includegraphics[width=0.45\linewidth]{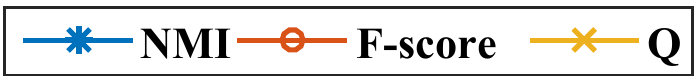}}
\caption{Parameter analysis: Effect of $K$ on two datasets with ground-truth community labels. } \label{fig:para1}
\vskip -0.1in
\end{figure}


\begin{figure}[!t]
\includegraphics[width=0.55\linewidth]{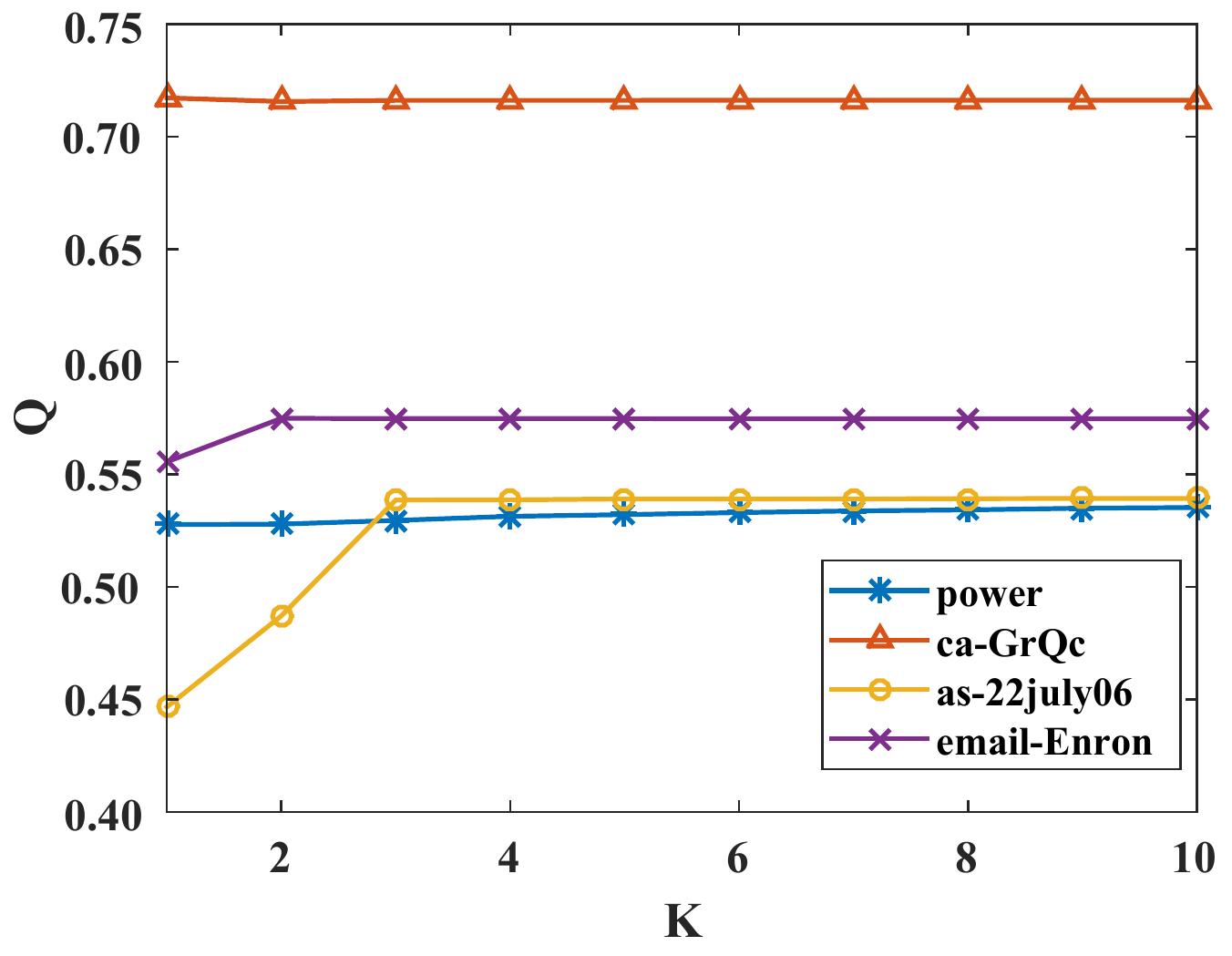}\vskip-0.1in
\caption{Parameter analysis: Effect of $K$ on four datasets without ground-truth community labels.} \label{fig:para2}
\vskip-0.1in
\end{figure}


In this section, parameter analysis is conducted to investigate the effect of the parameter $K$ on the performance of our \textbf{EdMot} method. Due to the space limit, we will take the Louvain method as the input graph partitioning method. However, similar analysis can be conducted by using any other graph partitioning methods.



The effect of $K$ on the performance of EdMot-Louvain is shown in~\figurename~\ref{fig:para1} and~\figurename~\ref{fig:para2}. Specifically, the effect of $K$ on polblogs and Cora is presented in~\figurename~\ref{fig:para1}. Since the motif-based hypergraphs of polbooks and email-Eu-core contain only one connected component, the discussion for these two datasets is omitted here. As can be seen, the scores of the evaluation measures hardly change as $K$ increases. This is because the motif-based hypergraph often contains a largest connected component that may consist of nodes from different communities and a large number of peripheral nodes that are isolated (see~\figurename~\ref{fig:corawm4} for illustration). Therefore, it is beneficial to partition the largest connected component in the hypergraph ($K=1$) into modules. 
Similar phenomenon can also be observed from the power and ca-GrQc datasets as shown in~\figurename~\ref{fig:para2}. As for the as-22july06 ($n=22963$, $m=48436$) and the email-Enron ($n=36692$, $m=367662$) datasets, which are relatively larger than other datasets, the best performance is achieved when $K=3$ and $K=2$ respectively. Thus, we set $K=3$ for the as-22july06 dataset, $K=2$ for the email-Enron dataset and $K=1$ for the other datasets in the experiments.



\section{Related Work}
\label{sec:relatedwork}
\subsection{Lower-Order Community Detection}
\label{sec:lowerorderCD}
Lower-order community detection methods discover communities by mainly leveraging the lower-order connectivity patterns of the network at the level of individual nodes and edges. For example, the Louvain method was proposed to reveal the hierarchical community structure by heuristically optimizing modularity~\cite{Vincent2008Fast}. The Nonnegative Matrix Factorization (NMF) method factorizes the node adjacency matrix into two nonnegative matrices and yields a new parts-based data representation~\cite{cai2008non,cai2011graph}. From the perspective of clustering nodes in the network into clusters, the Affinity Propagation (AP) clustering method was proposed to cluster nodes based on pair-wise similarities~\cite{frey2007clustering}. Similarly, the Spectral Clustering (SC) method was proposed to cluster nodes using eigenvectors of matrix~\cite{ng2002spectral}.
Besides, a generative model termed Stochastic BlockModel (SBM) was proposed to detect communities by fitting blockmodels to the network data~\cite{karrer2011stochastic,peixoto2014efficient}. And the permanence based method was also proposed to provide a more fine-grained view of the modular structure of the network~\cite{chakraborty2014permanence}. In addition, the label propagation based methods were developed, which possess some advantages such as the simplicity and nearly linear time complexity~\cite{raghavan2007near}.


\subsection{Motif-based Higher-Order Community Detection}
\label{sec:higherorderCD}

Network motifs are defined as patterns of interconnections occurring in networks at numbers that are significantly higher than those in the random networks~\cite{milo2002network,benson2016higher}. As the building blocks of the network, motifs are widely applied to unravel the design principles of gene regulation networks~\cite{shen2002network} and the underlying mechanisms of social networks~\cite{holland1977dynamic,wasserman1994social}.


Different from the lower-order community detection methods, higher-order community detection methods discover communities by leveraging the higher-order connectivity patterns of the network at the level of small network subgraphs, e.g., motifs.
For example, motif was used to define communities by extending the mathematical expression of Newman Girvan modularity in~\cite{arenas2008motif}. A graph sparsification principle based on graph motifs was proposed so as to improve the efficiency and quality for graph partitioning~\cite{zhao2015gsparsify}. The motif conductance based framework was also proposed to reveal higher-order organization of complex networks~\cite{benson2016higher}. Besides, a highly effective heuristic method was proposed based on motifs, where a random walk interpretation of the graph reweighing scheme was developed~\cite{tsourakakis2017scalable}. In addition, motifs have been leveraged in local higher-order graph partitioning~\cite{yin2017local}.

Despite the success in preserving the building blocks, the existing higher-order community detection methods may violate the original lower-order structure of the network. Specifically, the motif-based hypergraph may consist of a large number of connected components with various sizes and isolated nodes having no connections to any other nodes. In this case, they may suffer seriously from the hypergraph fragmentation issue. To address this issue, we propose an edge enhancement approach for motif-aware community detection (\textbf{EdMot}), which can not only leverage higher-order connections of the network but also overcome the hypergraph fragmentation issue.


\section{Conclusion}
\label{sec:conclusion}

In this paper, we for the first time propose a novel motif-aware community detection method termed \textbf{EdMot} for addressing the hypergraph fragmentation issue. Different from the existing higher-order community detection methods that directly operate on the possibly fragmented hypergraph, \textbf{EdMot} partitions the top $K$ largest connected components in the hypergraph into modules. And then, an edge enhancement approach is designed for enhancing the connectivity structure of the original network as follows. 1) First, a new edge set is constructed to derive a clique from each module. 2) Based on the new edge set, the original connectivity structure of the input network is enhanced to generate a rewired network, whereby the motif-based higher-order structure is leveraged and the hypergraph fragmentation issue is well addressed. After the edge enhancement, the rewired network is partitioned to obtain the higher-order community structure. Extensive experiments have been conducted to show the effectiveness of the proposed method.

\section*{Acknowledgments}
This work was supported by NSFC (61876193), National Key Research and Development Program of China (2016YFB1001003), Guangdong Natural Science Funds for Distinguished Young Scholar (2016A030306014), and Key Areas Research and Development Program of Guangdong (2018B010109007).

\bibliographystyle{ACM-Reference-Format}
\bibliography{myref}

\end{document}